\documentclass[twocolumn,showpacs,preprintnumbers,amsmath,amssymb]{revtex4}

\usepackage{graphicx}
\usepackage{dcolumn}
\usepackage{bm}

\input{psfig}
\begin{document}
\draft
\title{A Search for Sub-TeV Gamma Rays in Coincidence with Gamma Ray Bursts}
\author{J. Poirier, C. D'Andrea, P. C. Fragile,  J. Gress, G. J. Mathews,
 and  D. Race}
\address{University of Notre Dame, Center for Astrophysics, 
Department of Physics, Notre Dame, IN 46556}

\date{\today}

\begin{abstract}
We report on  a study of sub-TeV ($E_\gamma > 10$ GeV)
gamma-ray-induced muon secondaries
in coincidence with BATSE gamma ray bursts (GRBs).
Each TeV gamma ray striking the atmosphere 
 produces $\approx$0.2 muons whose identity and 
angle can be measured by the Project GRAND array.  
Eight GRB candidates were studied; seven were selected 
based upon the optimum product of (detected BATSE fluence) $\times$ (GRAND's acceptance). 
One candidate was added because it was reported as a 
possible detection by the Milagrito collaboration.
Seven candidates show a positive, though not statistically significant, 
muon excess.  The only significant possible coincidence shows an excess  of $466 \pm 171$ 
muons during the BATSE T90 time interval for GRB 971110.  
The chance probability of such an excess in GRAND's background at the
time of this event is $3 \times 10^{-3}$.  The chance probability of 
observing such an excess in one of the eight bursts studied here is 0.025.
If this event is real, 
the implied fluence of energetic ($> 10$ GeV) gamma rays necessary to account for 
the observed muon excess would require that most of the 
GRB fluence arrived in the form of energetic gamma rays.

\end{abstract}
\pacs{PACS Numbers: 95.85.Ry, 98.70.Rz, 98.70.Sa, 14.60.Ef, 95.55.Vj}
 
\maketitle

\section{Introduction}
     The mystery of the astrophysical origin for gamma ray bursts (GRBs) 
has been with us  for some time. 
As  of yet there is no consensus explanation for them.
Nevertheless, a likely scenario is a burst environment involving 
collisions  of an ultra relativistic $e^+-e^-$ plasma fireball 
\cite{Paczynski,Goodman,Sari}.  
These fireballs may produce low-energy gamma rays either by ``internal'' 
collisions of multiple shocks \cite{Paczynski94,Rees}, 
or by ``external'' collisions of a single shock with ambient clumps of 
interstellar material \cite{Meszaros}. 

In either of these possible paradigms, however,  it seems
likely that energetic ($\sim$TeV) gamma rays and/or neutrinos might 
also be produced along with the low energy GRB.
For example, inverse Compton scattering of ambient photons with
 relativistic electrons could produce high-energy gamma rays \cite{fragile}.
Alternatively, baryons would be accelerated along with
the pair plasma to very high energies \cite{Waxman95,Waxman97,totani}.  
Synchrotron emission from energetic protons 
\cite{totani},
or hadroproduction of pions in the burst environment\cite{Waxman97} 
and subsequent $\pi^0$ gamma ray decay might also yield 
a spectrum of energetic gamma rays.

Thus, it is plausible that energetic gamma rays 
could be emitted in coincidence with a lower-energy GRB.  
These energetic gamma rays would, however, be attenuated by the 
infrared background.  Hence, their detection will require that the 
source be nearby and/or that their flux constitute a significant
fraction of the energetic output of the source.  
Nevertheless, such energetic gamma rays, if detected, could provide 
valuable clues as to the baryon loading,
Lorentz factor, and ambient magnetic field
 of the relativistic fireball. They might also provide
a means to distinguish between 
an internal vs. external shock origin for the bursts.  
This paper, therefore,  reports on an independent search for 
the possible coincidence of high-energy gamma rays with GRBs.

There presently exists at least some evidence
for a possible association of energetic gamma rays with low-energy GRBs 
in previous literature.
EGRET detected seven GRBs which emitted high energy photons in the
$\sim 100$ MeV to 18 GeV range 
\cite{Schneid,Hurley,Catelli}.  
There have also been some results from the Tibet air shower array 
suggestive of 
gamma rays beyond the TeV range \cite{Amenomori,Padilla}, although
these results were not claimed as a firm detection.  There
has also been reported evidence for TeV emission in one burst
out of 54 BATSE GRBs in the field of view of the Milagrito
detector \cite{milagro}. 
The chance probability for that  event 
(out of 54 trials) was found to be $1.5 \times 10^{-3}$.
During the present study, this  Milagrito 
event was also within
the field of view of GRAND, but at a relatively low elevation.

In this paper we report a marginal ($2.7 \sigma$) detection of 
energetic ($E_\gamma > 10$ GeV) gamma rays in coincidence with 
one BATSE GRB (971110).  The chance probability of such an excess
from this burst is $3 \times 10^{-3}$.  The chance probability of
such an excess from our sample of eight bursts is 0.025.  
The constraint that this detection might place on the
spectrum of energetic photons from the burst environment is discussed.
The analysis of GRAND's data for the GRB reported as a possible 
detection by Milagrito finds an insignificant excess of muons 
($1.2 \sigma$, including statistical and systematic errors).  

\section{Project GRAND}

GRAND is located just north of the University of Notre Dame 
campus, approximately 150 km east of Chicago and 220 m above sea level at 
86.2$^o$ W and 41.7$^o$ N.   
It detects cosmic ray secondaries at ground level by means of 64 
tracking stations of proportional wire chambers (PWCs)\cite{Gress90}.  
Each station has four PWC 
detectors; each detector contains an x-plane (x=eastward) and a 
y-plane (y=northward) oriented horizontally and 
stacked vertically (z) \cite{Linsley,Poirier}.  
The planes have an active area of 1.29 m$^2$ comprised  of
80 detection cells (each 14 mm $\times$ 19 mm $\times$ 1.1 m).
Each secondary muon is measured to 0.26$^o$ absolute precision 
(average value in each of two orthogonal planes) \cite{Gress91}.  
At present, GRAND's total detector area is 83 m$^2$; earlier data 
had smaller areas.
A 51 mm thick steel plate inserted between the third and fourth PWCs allows 
muon tracks to be distinguished from electrons;
96\% of muon tracks are correctly identified as muon tracks.
Since, for single tracks, only 1/4 are electrons and 4\% of these 
electrons are misidentified as muons, the muon sample has only 1\%
contamination from electrons.  
The data presented here are from single-track triggers which ignore time 
coincidences between stations; only muon candidates are selected.

GRAND utilizes the fact that gamma rays have a detectable signal 
of muons from gamma-hadro production in the atmosphere.  
The pions thus produced subsequently decay to muons which can
reach detection level making it possible to study coincidences between
GRBs and gamma ray showers in the $E_\gamma \ge$ 10 GeV 
energy region.  
This energy threshold (10 GeV) depends slightly 
upon the spectral index of the gamma ray spectrum (see 
Figure 6 in Ref.~\cite{ajs}) and is not a sharp threshold.    
It has been estimated \cite{Mannheim} that 
the GRB rate for a threshold 
energy larger than 200 GeV is $\sim $10 GRBs per year; 
for the $\sim $10 GeV threshold energy of GRAND, 
this rate should be even greater.

Since the study reported here is similar to and includes the
event reported  by the Milagrito collaboration,
we note the differences between GRAND and Milagrito.
Whereas GRAND detects secondary air-shower muons with PWCs, Milagrito
detects secondary air shower charged particles by Cerenkov light
as the shower particles traverse a light tight
water reservoir.   The main differences between the two detectors
are that Milagrito has a larger active detection area, a 
somewhat better angular resolution in the single-track mode of interest here
($\sim 1^o$ for Milagrito vs. $\sim 5^o$ for GRAND), and a higher
detection threshold ($\sim 1$ TeV for Milagrito vs. $\sim 10$ GeV for GRAND).
The lower detection threshold for GRAND implies that it is  sensitive
to a lower energy part of the primary gamma ray spectrum which is not as likely 
to have been extinguished by internal \cite{Waxman97} or
intergalactic absorption \cite{Salamon,totani00}. 
Thus, even though no individual detection was overwhelmingly significant,
the fact that a slightly positive signal to background was
deduced for all but one of the eight candidate bursts investigated
might suggest that 
$\ge$10 GeV emission in association with low-energy GRBs 
is not altogether uncommon.  

GRAND is a unique detector facility which measures the angles of single 
tracks and identifies which are muons.  Inferring the implied gamma ray 
flux from the detected muons, however, requires confidence in
the ability to simulate the muon production 
from gamma hadroproduction and muon propagation in the
atmosphere.  Recently, we have made a detailed analysis \cite{ajs,Fasso} of the
spatial and energy distribution of muons in $\gamma$-induced air showers.
These simulations are based upon the FLUKA Monte-Carlo (MC) code.
Unlike most MC codes used in cosmic ray research,  this
simulation is not specialized
for this particular field but is a multipurpose particle transport
code which has been tested in many diverse applications such
as proton and electron accelerator shielding, calorimetry, medical
physics, etc.  It has therefore been verified against a large amount
of nuclear experimental data and indirectly validated by comparisons
with shower measurements obtained both at accelerators and 
in cosmic ray experiments.  Of particular relevance to the present study
is the fact that this code has been shown \cite{Battistioni}
to accurately predict hadron-generated muon spectra at different heights
in the atmosphere.  Hence, we expect its application here to photon-generated
muon spectra be good to
the $\sim$few \% statistical accuracy of the simulations.

The MC  simulation  of \cite{Fasso} shows that 
a 1 TeV gamma ray normally incident upon the 
earth's atmosphere produces an average of 0.23 muons which reach GRAND.  
GRAND thus paradoxically uses muons as a $signal$ for gamma ray primaries.  
In the energy region $\ge$ 10 GeV, the muon statistics 
are quite high;
the current  all-sky rate for recording identified muons is about $2400$~Hz 
or 8.6 million muons per hour.

     These secondary muons are primarily the result of interactions of 
primary cosmic rays with the atmosphere producing pions, which then decay to 
muons.  
The pions are produced at small angles relative to the primary cosmic rays; 
the non-interacting pions decay to muons which emerge at small angles relative to the pion;  
the muons are then deflected in the earth's magnetic field and scattered 
in the atmosphere resulting in an effective net angular resolution
of about $\pm ~5^o$ (for the primary cosmic ray 
in each of the two orthogonal directions; this resolution depends slightly 
upon the primary energies or spectral index).  
The muon threshold detection energy is 0.1 GeV for vertical tracks to 
penetrate the 50 mm steel plate; however, 
these muons must have been born with  at least 
several GeV of energy to penetrate the 
atmosphere in order to reach the detectors.  

The precise response to primary gamma rays is described
with a Monte Carlo calculation 
(see Figure 6 of  \cite{prf}); 
the result depends upon the assumed spectral index.  
The primary gamma-ray spectrum is assumed to be  of the form $E_\gamma^{~\beta}$ 
with a spectral index $\beta = -2.41$ (the average of the 
spectral indices reported in the third EGRET catalog 
\cite{Hartman}). This spectrum 
of primary gamma rays is then multiplied by the number of muons per gamma ray
 which reach detection 
level \cite{Fasso,prf}.  This determines the number of detectable muons
as a function of the primary gamma-ray energy.  Qualitatively, each 
primary gamma ray  produces a number of muons.  This number at first  increases
sharply for gamma-ray primary energies from 1.5 to 10 GeV 
and then falls slowly for energies above 10 GeV.  For a softer spectral 
index, the muon response peaks at a slightly lower primary gamma-ray 
energy and then falls off more rapidly above 10 GeV.  For convenience, we 
characterise this response shape as having a threshold of $\sim$10 GeV 
for the primary gamma-ray spectral indices of interest.  

GRAND's ability to correlate short bursts of muons with an identifiable 
source of primary radiation has been shown in a detection which 
was coincident with a solar flare on 15 April 2001.  The 
statistical significance of this observation was at the level of 
6 $\sigma$ for a ground level event of 0.6 hours duration \cite{gle}.  

\section{Data Analysis}
     To analyze GRB coincidences 
the complete GRB table, the flux table, and the duration table
were downloaded from  the BATSE archive \cite{BATSE}.
Two of the candidate GRBs are listed in the BATSE 4b catalog, the other
more recent events are in the BATSE archives.  
Table 1 lists some of the BATSE data including: the date of the trigger (GRB), 
the trigger number (Trig); 
the time duration for 90\% of the burst's counts to occur (T90) 
in seconds, 
the right ascension (RA), declination (Dec), and  
the BATSE angular error ($\delta\theta$), all in degrees.  
Next to these BATSE entries are 
the calculated angular elevation (in degrees) 
above GRAND's horizon (denoted Elev) and  our selection criterion
($LogLk$) described below.
The last three columns of Table 1 summarize 
the muon secondaries observed by GRAND
within a $\pm 5^o$ square angular window
centered on the burst location during the BATSE T90 time interval.  

\subsection{Selection Criterion}
For each GRB, an approximate  total acceptance factor, A, was calculated 
for the detector stations of GRAND:
\begin{equation}  
A \simeq \bigl[(1 - g\times\tan{\theta_x}) (1 - g\times\tan{\theta_y})
\times\cos{\theta}\bigr]\times\cos^2{\theta}
\end{equation}
where the geometrical factor $g \equiv  d/L$, with 
$d$ the vertical spacing between the top and bottom plane 
(0.61 m), and $L$ the length of a single plane (1.1 m). 
The angles $\theta_x$ 
and $\theta_y$ are the angle 
of the track from vertical ($\theta$, the complement of Elev) 
projected onto the xz- and yz-planes, respectively.  
The quantity in square brackets is the geometrical acceptance 
of a detector station; 
this acceptance has been
multiplied by $\cos^2{\theta}$ to account for the added absorption 
of a muon traversing an increased path length through 
the atmosphere for tracks inclined from the vertical.  

As a rudimentary criterion to select which events to analyze,
the relative likelihood $LogLk$ of GRAND to observe each GRB,  is 
taken as proportional to the (base 10) logarithm of the 
product of the total acceptance factor, equation (1), times the  BATSE 
fluence in the highest of four energy bins (i.e. $E_\gamma \ge$ 300 KeV).   
The GRBs with the 21 largest values for  $LogLk > 3$
from the BATSE archive were selected for 
further consideration.   In addition, it was recognized
that the  Milagrito  
event was in the field of view of
GRAND, and even though it did not satisfy the 
selection criterion, it was included in the analyses.  

The entries in Table 1  are ordered 
with the highest value of $LogLk$ at the top.  
This $LogLk$ factor, however,  did not take into account the fact that the 
array was under continuous construction during this time and 
had varying (usually increasing) numbers of operational detectors 
at a given time.
The data for each GRB analyzed were checked to ensure no huts were 
turning on or off during the time of analysis for that GRB.  
Data for 11 of the top 21 GRBs by this $LogLk$ criterion 
were available on archived data tapes.  
Of the  11, one was found to have individual stations with large time 
dependent inefficiencies and was discarded.  
In addition, three of the candidate 
GRBs were before 1994 when the detector area was small 
and had errors in the clock's seconds bit.
Since these three bursts had the shortest T90 times, they were 
most sensitive to the precise time.  Furthermore, the
BATSE angular errors were largest (two were comparable to the 
total width of our analysis window).  Therefore, these three events
were eliminated from the present analyses.  
[These three events were included in a 
previous preliminary analysis \cite{icrc} 
with chance probabilities of +1.6, -0.1, and -1.4$\sigma$ (Stat.)]  
The calculation of time from the data was checked 
to ensure that the times of the remaining GRBs are correct.

 In a preceding preliminary analysis \cite{icrc}, 
we employed an angular window using fixed 
angles of right ascension (RA) and declination (Dec).
However, the signal and background data in this analysis showed 
systematic variations due to:  
1) The local angles $\theta_x$ and $\theta_y$ change with time 
during a burst, which  means that event tracks are
recorded with differing numbers of wire combinations 
leading to systematic fluctuations in the count rate.   
2) The detector's solid angle increases as 
a fixed RA and Dec moves toward the vertical (or conversely).  
3) There is dead time (tens of milliseconds) 
whenever: a) the data input is stopped in order to write an event buffer to 
magnetic tape; or 
b) a trigger is received for a concurrently running air-shower experiment.
These systematic effects 
were accounted for in the current analysis method as follows:

First, the BATSE GRB locations in RA and Dec were transformed to a 
horizon coordinate system of elevation and azimuth and then 
projected onto the xz-plane ($\theta_x$) and the yz-plane ($\theta_y$).
A window of $\pm5^o$ in $\Delta \theta_x$ and $\Delta \theta_y$
was centered on the location of the GRB. 

As time progressed during the T90 burst interval, 
the local windows of $\theta_x$ and $\theta_y$ stay fixed 
thus removing problems (1) and (2) above.  
To correct for the dead time (3), 
the total event rate over the whole sky was employed 
as a high-statistics measure of the live time of each time bin; 
each bin's data were corrected for its corresponding live time.  
After this correction, the data inside the 
angular cuts were almost independent of the random dead times.
These corrected numbers of muons inside the angular window were studied 
to obtain the (background + signal) within the T90 interval.  
The background was determined  during a time interval of $\approx$20$\times$T90 
before the start of the BATSE trigger
(except in the case of GRB 971110 it was $\approx$10$\times$T90 
as the longer interval of time was not available on the same data 
tape for this event).

\subsection{Significance}
The signal (Sig) calculated for a GRB is the difference between the total 
counts $N_{on}$  inside the T90 interval (corrected for dead time) and 
the background counts  $N_{off}$ normalized to the 
live time of the T90 time interval and corrected for dead time.  
The statistical significance  $Sig/\delta Sig$ (number of standard deviations
above background) of each event was determined according to the 
likelihood ratio method
of Li and Ma \cite{lima}, 
\begin{eqnarray}
Sig/\delta Sig &=& \sqrt{2} \biggl\{ N_{on} ln{\biggl[ {1 + \lambda\over \lambda}
\biggl({N_{on} \over N_{on} + N_{off}} \biggr) \biggr]} \nonumber \\
&& + N_{off} ln{\biggl[(1 + \lambda)
 \biggl({N_{off} \over N_{on} + N_{off}} \biggr) \biggr]} \biggr\}^{1/2}~~,
\end{eqnarray}
where $\lambda$ is the ratio $\lambda \equiv t_{on} /t_{off}$.  
This method makes a best accounting of the true significance of an event
when the background and event time intervals differ, i.e. $\lambda \ne 1$.
For all of these events
$ \lambda \approx  0.05$ except for GRB 971110 for which $\lambda \approx  0.10$. 
The statistical errors from this analysis are listed in the last column of  Table 1.

The data for these GRBs were tested 
before, during, and after the signal region to see 
if any station's data 
had excessive noise or erratic time dependences which could 
erroneously mimic a GRB signal.  
These rates exhibited no pathological time structure.  

\begin{figure}[htb]

\vskip .2 in

\mbox{\psfig{figure=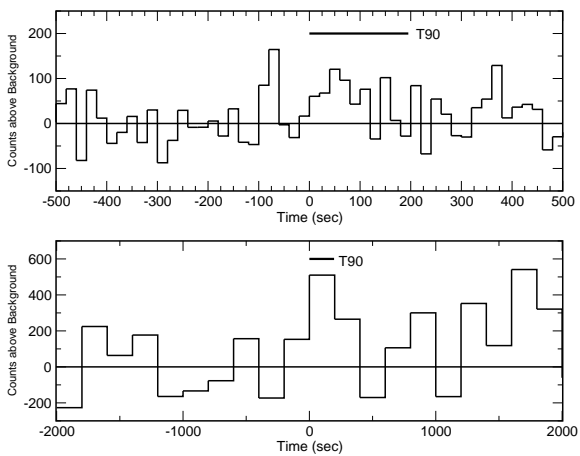,width=3.4in,angle=270}}
\caption{Background-subtraced event rate in the Project GRAND
array before and after the BATSE 6472 event trigger.  The T90
interval for this burst is indicated by  a horizontal line
 above each of the two histograms.}
\label{ltcrv}
\end{figure}

The only possibly significant single observation among the eight 
GRBs in Table 1 
corresponds to GRB 971110.  
The time distribution of the 
data before and after the BATSE event trigger are illustrated in
Figure \ref{ltcrv}.  Here the data are plotted  in a manner similar 
to the way in which the Milagrito 
\cite{milagro} event was presented,
i.e., with two different values for the time-bin resolution and 
for different durations before and after the GRB.
In this figure negative background-subtracted events are 
also included giving added information on the fluctuations in 
the background.
There are excess events in the T90 window (and possibly
before and after as well).   The muon excess in the T90
window  is $3.3 \sigma$ 
above the $\sqrt{N}$ fluctuations expected from pure counting statistics.
Nevertheless, it is also possible that the background fluctuates 
in excess of the expected  $\sqrt{N}$ statistics. This could happen  
due to real variations in the muon arrival rate at a given angle or to
variations in the detector response not already accounted for in our dead time 
correction.  In either case, the additional systematic error associated
with such possible excess fluctuations can be deduced directly from the observed data rate
as we now describe.

\subsection{Systematic Error}
As a check on systematic errors in the signal for GRB 971110, 
similar angular sections of the sky (which have the same 
absolute values of $\theta_x$ and $\theta_y$ 
and thus the same average counting rates) 
and the same time interval (T90) but at different, neighboring times were analyzed.    
These time intervals span $\pm$26~hours 
relative to the BATSE trigger for GRB 971110.
The data on the tape were also analyzed 
 for T90 intervals beginning at times 
before and after the GRB, and 
for different  $\theta_x$ and $\theta_y$ locations in the sky but at
the same elevation angle.  
This analysis 
was done in 
the same way as described in the preceding paragraphs.  

\begin{figure}
\mbox{\psfig{figure=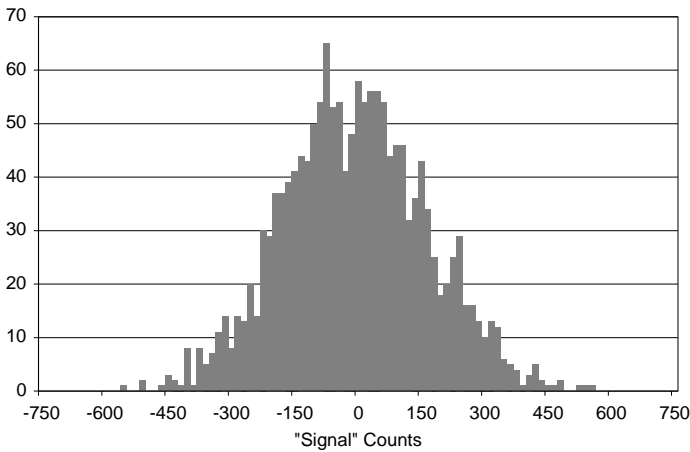,width=3.6in,angle=0}}
\caption{Distribution of fluctuations above and below background
for  1587 random T90-intervals
observed by GRAND near the time of GRB 971110}
\label{sigdist}
\end{figure}

Figure \ref{sigdist}  shows the distribution of fluctuations
above and below background  for 1587 
separate analyses (not including the time and angle of 
the GRB 97110 event).
The resulting distribution is 
centered on zero, and is approximately Gaussian in shape.
However, this distribution
has a larger width than that calculated from the Poisson  random 
statistics of the counts within the T90 and background intervals.  
The standard deviation width of this distribution is 171 counts, 
whereas the expected
statistical deviation is only 141 counts based upon the
background count rate.  A total
statistical plus systematic standard deviation is 171 counts 
for this distribution. 
Adopting this  total error as the quadrature sum of the statistical
and systematic errors, then 97 counts are ascribed  
as the additional systematic error in our analysis for GRB 971110.
With this additional systematic error, the ratio of signal-to-noise
becomes $Sig/\delta Sig = 2.72$.

The probability of a $+2.7 \sigma$ fluctuation in a Gaussian white-noise
distribution is $4.9 \times 10^{-4}$.  
However, the histogram in Figure \ref{sigdist} can be used to measure the 
probability of GRAND's detection being real without 
assuming a distribution which is Gaussian.  
In this histogram there are 10 random
fluctuations either $\ge$ +466 or $\le$ -466.  
This corresponds to a probability of $3 \times 10^{-3}$ for a 
background fluctuation to produce a $\ge$ +466 signal. 
For one event out of the  8 analyzed
to exhibit this much deviation would
correspond to a random probability of 0.025. Thus, the statistics of
this event are interesting but not compelling.  

The total statistical plus systematic error
derived above  is consistent with a white-noise fluctuation that 
would scale with the statistical error.  In this case,  
it is reasonable to take the ratio of total to statistical 
error as a constant  independent of the particular T90 interval
in question.
This constant factor ($\sim 1.2$) 
can then be used to roughly estimate the systematic
 error for the remaining events.  These remaining events (except one) 
indicate positive (though not significant) deviations.


\section{Discussion}

Accepting for the moment the proposition that the muon excess associated
with GRB 971110 may be real, it is worthwhile to consider some of the implications
of this event. First, we wish to consider the possibility of whether these events
could be explained by a naive extension of the low-energy burst spectrum.
Indeed, it is known from EGRET detections \cite{Schneid} that it is possible for
the spectrum of the low-energy burst detected by BATSE to  extend up to
$\sim$ 10 GeV with no change in the spectral index. On the other hand,
such an extension is not guaranteed.  For example, no sub MeV photons
were observed at the time that EGRET observed an 18 GeV gamma ray.

To explore the possibility that the observed muon excess
could be explained by a straightforward extension of the BATSE spectrum,  
the  GRB 971110 spectrum
was fit using  the standard BATSE empirical spectral 
model \cite{band}: 
\begin{eqnarray}
{d^3N\over dEdAdt} &=& B\left({E \over
\mbox{100 KeV}}\right)^\alpha \exp\left(-{E(2+\alpha) \over
E_{peak}}\right)
\end{eqnarray}
for low energy, $E < (\alpha-\beta)E_{peak}/(2+\alpha)$, and
\begin{eqnarray}
 = B\left\{{(\alpha-\beta)E_{peak} \over [\mbox{100
KeV}(2+\alpha)]}\right\}^{\alpha-\beta} \left({E \over \mbox{100 
KeV}}\right)^\beta \exp(\beta-\alpha)
\end{eqnarray}
for $E \ge (\alpha-\beta)E_{peak}/(2+\alpha)$.
The best-fit parameters for this burst are $\alpha=-1.02 \pm 0.04$, 
$\beta=-2.33 \pm 0.11$, 
$B=0.0095 \pm 0.0003$, and $E_{peak}=303 \pm 17$ KeV, with a 
reduced $\chi^2$ of 0.93.  

\begin{figure}
\mbox{\psfig{figure=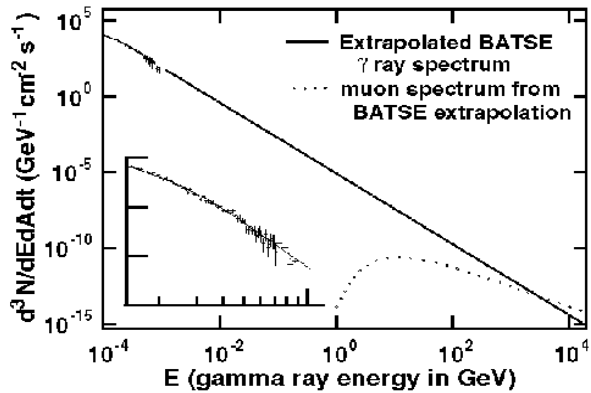,width=3.3in,height=5.5 cm}}

\vspace{.07in}       
\caption{
Extrapolation of the best-fit energy spectrum for BATSE GRB 971110.  
A portion of the BATSE data are shown in the inset.  
Multiplying the extrapolated BATSE gamma ray spectrum 
by the gamma 
to muon conversion efficiency (as calculated by FLUKA \cite{Fasso}) 
gives the spectrum of muons which would be observed by 
GRAND.
}
\label{spectrum}
\end{figure}

The solid line in Figure \ref{spectrum}
 shows the optimum fit to the 
BATSE spectrum extended to the energy range of GRAND.
Note that this naive extension is probably an over estimate, since absorption
due to pair production from interactions of gamma rays 
with the intergalactic infrared background is 
expected to become significant above $\sim$ 200 GeV \cite{Salamon,totani00}.
 
This extended spectrum was folded with the known efficiency, 
$\epsilon_\mu$, of high-energy photon to muon conversion in the 
atmosphere (in the region of GRAND 
$\epsilon_\mu \approx 0.23 E^{1.17}_{TeV}$ \cite{Fasso})
to yield the spectrum of muons at GRAND 
expected from this extrapolation of the BATSE spectrum 
(dotted curve in Figure \ref{spectrum}).  The broad peak of the muon 
spectrum at around 10 GeV illustrates the peak (threshold) sensitivity
of GRAND for this particular gamma ray spectrum.
Integrating this muon spectrum over the primary gamma ray energy, $E_\gamma$, 
yields the number of muons per area per time based upon this extrapolation.  
Multiplying this result by the effective muon detection area of GRAND 
at the time of this burst and by its T90 time interval 
yields 0.3 $\pm$ 0.6 muons--well below the 
observed $466 \pm 171$ excess muons detected for this event.  
Even if all of the fit variables are adjusted to their respective upper 
1$\sigma$ limits, only four muons would be expected.  

Clearly, a simple, naive (very long) extrapolation from the BATSE
spectrum is inconsistent with the observed muon excess at GRAND.  
In a related paper
\cite{fragile} the specific constraints which
these data place upon possible mechanisms for
energetic gamma ray emission
such as energetic proton-synchrotron emission
\cite{totani}, inverse-Compton scattering from relativistic electrons,
 or hadronic production of pions  in the burst \cite{Waxman97} are discussed.
 
\section{ Conclusion}

We have completed a search for evidence for energetic  sub-TeV gamma rays
in coincidence with low-energy gamma ray bursts.
No convincing evidence is found, though there
is a possible $2.7 \sigma$ detection associated with the
best candidate burst.  There is an
insignificant,  $1.2 \sigma$, muon excess 
observed in association with the Milagrito event (GRB 970417a), 
as well as a slight positive excess for
all (but one) of the remaining events investigated.

Based upon an analysis of the most significant event, 
we conclude that if the detected muon excess is real, 
then  probably a new sub-TeV component to the gamma ray spectrum is required.
A determination of the magnitude of this component, however, will require 
knowledge of the source spectrum as well as the effects of  photon absorption
both in the burst environment and enroute. 
Studies along this line \cite{fragile} imply
that most of the GRB emission would be in sub-TeV gamma-rays.

\acknowledgments
The authors wish to acknowledge contributions of
D. Baker, J. Carpenter, S. Desch, C.F. Lin and A. Roesch in the analysis; 
this research has made use of data obtained from the High Energy
Astrophysics Science Archive Research Center (HEASARC),
provided by NASA's Goddard Space Flight Center; 
thanks to M. Briggs for his help in fitting the 
BATSE spectra; and to T. Totani for help with the intergalactic 
absorption calculations.  
Project GRAND's research is presently being funded through a grant from the 
University of Notre Dame and private grants.   One of the authors (GJM)
wishes to acknowledge support from DoE Nuclear Theory Grant DE-FG02-95-ER40934.

\begin{table*}
\caption{Summary of Events Analyzed.}
\label{grbtable}
\begin{ruledtabular}
\begin{tabular}{lccccccccccccc}
GRB&Trig&T90&RA&Dec&$\delta \theta$&Elev& $LogLk$ & $N_{on}$ & $\lambda N_{off}$ & $N_\mu \pm \sigma_{Tot}$ &
$\sigma_{Stat}$   \\
\hline
971110 & 6472 & 195.2 & 242 & 50 & 0.6   & 81 & 5.18 & 18,286 & 17,820  & $466 \pm 171 $ &141    \\
990123 & 7343 &  62.5 & 229 & 42 & 0.4   & 56 & 5.13 & 1,079  & 1,076   & $3   \pm 36  $ &30    \\
940526 & 2994 & 48.6  & 132 & 34 & 1.7   & 66 & 4.68 & 498    & 478     & $20  \pm 28  $ &23    \\
980420 & 6694 & 39.9  & 293 & 27 & 0.6   & 68 & 4.02 & 1,456  & 1,417   & $39  \pm 47  $ &39    \\
960428 & 5450 & 172.2 & 304 & 35 & 1.0   & 70 & 3.83 & 3,990  & 3.933   & $57  \pm 78  $ &64    \\
980105 & 6560 & 36.8  & 37  & 52 & 1.4   & 79 & 3.46 & 2,214  & 2,229   & $-15 \pm 61  $ &50    \\
980301 & 6619 & 36.0  & 148 & 35 & 1.3   & 76 & 3.17 & 2,053  & 2,007   & $38  \pm 56  $ &46    \\
970417a& 6188 & 7.9   & 290 & 54 & 1.6   & 62 & 2.08 & 186      & 166   & $20  \pm 17  $ &14    \\
\hline
\end{tabular}
\end{ruledtabular}
NOTE: T90 is in seconds. Angles RA, Dec, $\delta\theta$, and Elev are in degrees.
\end{table*}

\begin{references}
\bibitem{Paczynski}B. Paczy\'nski, Astrophys. J. {\bf 308}, L43 (1986).
\bibitem{Goodman}J. Goodman,  Astrophys. J. {\bf 308}, L47 (1986).
\bibitem{Sari}R. Sari, T. Piran, and R. Narayan, Astrophys. J. {\bf 497}, L17 (1998).
\bibitem{Paczynski94}B. Paczy\'nski and  G. Xu, Astrophys. J. {\bf 427},
708 (1994).
\bibitem{Rees}M. Rees and P. M\'esz\'aros, Mon. Not. R. Astron.
Soc. {\bf 258}, P41 (1992).
\bibitem{Meszaros}P. M\'esz\'aros and M. Rees, Mon. Not. R. Astron.
Soc. {\bf 269}, L41 (1994).
\bibitem{fragile}P. C. Fragile {\it et al.}, Submitted to Astrophys. J. (2002).
\bibitem{Waxman95}E. Waxman, Phys. Rev. Lett. {\bf 75}, 386 (1995).
\bibitem{totani}T. Totani, Astrophys. J. {\bf 502}, L13 (1998); 
{\bf 509}, L81 (1998).
\bibitem{Waxman97}E. Waxman and J. Bahcall, Phys. Rev. Lett. {\bf 78}, 2292 (1997).
\bibitem{Schneid}E. J. Schneid {\it et al.}, Astron. Astrophys. {\bf 255}, L13 (1992).
\bibitem{Hurley}K. Hurley, Nature {\bf 372}, 652 (1994).
\bibitem{Catelli}J. R. Catelli, B. L. Dingus, and  E. J. Schneid,
in {\it Gamma ray Bursts}, edited by C. A. Meegan,
AIP Conf. Proc. No. 428 (AIP, New York, 1997).
\bibitem{Amenomori}M. Amenomori {\it et al.}, Astron. Astrophys. {\bf 311}, 919 (1996).
\bibitem{Padilla}L. Padilla {\it et al.}, Astron. Astrophys. {\bf 337}, 43 (1998).
\bibitem{milagro} R. Atkins {\it et al.} (Milagrito Collaboration), Astrophys. J. Lett.,
{\bf 553}, L119 (2000); R. Atkins {\it et al.}, Submitted to Astrophys. J. 
(2002); I. R. Leonor {\it et al.} 
in {\it Proceedings of the 26th International Cosmic Ray Conference}, 
{\bf4}, 12, edited by D. Kieda, M. Salamon, and B. Dingus (Salt Lake City, 1999).
\bibitem{Gress90}J. Gress {\it et al.} in {\it Proceedings of the 21st 
International Cosmic Ray Conference}, {\bf10}, 335, 
edited by R. J. Protheroe (Adelaide, 1990).
\bibitem{Linsley}J. Linsley {\it et al.}, J. Phys. G {\bf 13}, L163 (1987).
\bibitem{Poirier}J. Poirier {\it et al.}, Nucl. Phys. B {\bf 14A}, 143 (1990).
\bibitem{Gress91}J. Gress {\it et al.}, Nucl. Instrum. Methods {\bf A302}, 368  (1991).
\bibitem{ajs}J. Poirier, S. Roesler, and A.\ Fass\`o, Astroparticle Physics {\bf 17}, 441 (2002).
\bibitem{Mannheim}K. Mannheim, D. Hartmann, and B. Funk, Astrophys. J. {\bf 467}, 532 (1996).
\bibitem{Salamon}M. H. Salamon and F. W. Stecker, Astrophys. J. {\bf 493}, 547 (1998).
\bibitem{totani00} T. Totani, Astrophys. J. Lett., {\bf 536}, L23 (2000).
\bibitem{Fasso}A.\ Fass\`o and J. Poirier, Phys. Rev. {\bf D63}, 036002 (2001).
\bibitem{Battistioni}S. Roesler, W. Heinrivh, and H. Schraube, Radiat. Res.,
{\bf 149}, 87 (1998); 
G. Battistioni et al., Astropart. Phys., {\bf 9}, 277 (1998);  
G. Battistioni et al., Astropart. Phys., {\bf 12}, 315 (2000).
\bibitem{prf}J. Poirier, S. Roesler, and A. Fass\`o, Astroparticle Phys. {\bf 17}, 441 (2002).
\bibitem{Hartman}R. C. Hartman et al., Astrophys. J. Suppl. Ser. {\bf 123}, 79 (1999).  
\bibitem{gle}J. Poirier and C. D'Andrea, 
 Journal of Geophysical Research, Space Physics (2002), in press.  
\bibitem{BATSE} 
http://www.batse.msfc.nasa.gov/batse/
\bibitem{icrc}T. F. Lin {\it et al.} in 
{\it Proceedings of the 26th International Cosmic Ray 
Conference}, {\bf4}, 24, edited by D. Kieda, M. Salamon, and B. Dingus 
(Salt Lake City, 1999).
\bibitem{lima}T.-P. Li and Y.-Q. Ma, Astrphys. J., {\bf 272}, 317 (1983).
\bibitem{band} D. Band {\it et al.}, Astrophys. J. {\bf 413}, 281 (1993).
\bibitem{fenimore} E. E. Fenimore and E. Ramirez-Ruiz, Astrophys. J. Submitted (2000), astro-ph/0004176;
D. E. Reichart, et al. Astrophys. J., {\bf 552}, 57 (2001).
\end{references}
\end{document}